\documentclass[reprint,twocolumn,superscriptaddress,showpacs,nofootinbib,notitlepage,preprintnumbers]{revtex4-2}

\usepackage{graphicx}
\usepackage{hyperref}
\usepackage[dvipsnames]{xcolor}
\usepackage{amsmath,amsthm,amssymb}
\usepackage{comment}
\usepackage{enumerate}
\usepackage{ulem}

\renewcommand{\Re}{\mathrm{Re}\,}

\DeclareMathOperator{\Tr}{Tr}

\begin{document}
\preprint{LA-UR-24-24672}
\preprint{INT-PUB-24-023}

\title{Quantum states from normalizing flows}

\author{Scott Lawrence}
\email{srlawrence@lanl.gov}
\affiliation{Department of Physics, University of Colorado, Boulder, CO 80309, USA}
\affiliation{Los Alamos National Laboratory Theoretical Division T-2, Los Alamos, NM 87545, USA}
\author{Arlee Shelby}
\email{ashelby@ncsu.edu}
\affiliation{Department of Physics, University of Colorado, Boulder, CO 80309, USA}
\affiliation{Department of Physics, North Carolina State University, Raleigh, NC 27695}
\author{Yukari Yamauchi}
\email{yyama122@uw.edu}
\affiliation{Institute for Nuclear Theory, University of Washington, Seattle, WA 98195, USA}

\date{\today}

\begin{abstract}
	We introduce an architecture for neural quantum states for many-body quantum-mechanical systems, based on normalizing flows. The use of normalizing flows enables efficient uncorrelated sampling of configurations from the probability distribution defined by the wavefunction, mitigating a major cost of using neural states in simulation. We demonstrate the use of this architecture for both ground-state preparation (for self-interacting particles in a harmonic trap) and real-time evolution (for one-dimensional tunneling). Finally, we detail a procedure for obtaining rigorous estimates of the systematic error when using neural states to approximate quantum evolution.
\end{abstract}

\maketitle

\section{Introduction}\label{sec:introduction}

Quantum many-body physics is intrinsically difficult to study on classical computers. This is largely a result of the fact that representing the quantum state requires an exponential amount of classical information in the number of particles. Efficient classical simulations of quantum systems therefore rely on the ability to either avoid needing to represent a quantum state directly, as with the Monte Carlo methods underlying e.g.~lattice QCD, or the ability to work in a (typically exponentially) reduced subspace of the full Hilbert space.

A modern machine learning-based approach to the many-body problem may be found in neural quantum states, which take the latter approach. These were originally proposed as both a variational ansatz for ground-state physics and a way to probe quantum time-evolution~\cite{carleo2017solving}; see~\cite{Lange:2024nsr} for a recent review. In general, a neural quantum state refers to a neural-network parameterization of a subset of the Hilbert space of some physical system. The neural network can then be used as a variational ansatz for approximating the ground-state, to be optimized via efficient gradient descent. Additionally, the time-dependent Schr\"odinger's equation may be approximated in this restricted space, allowing access to real-time evolution. Since being proposed, neural quantum states have since been applied to a variety of systems, including spin liquids~\cite{li2021learning,PhysRevX.11.031034,zhang2017machine}, Fermi gasses~\cite{Kim:2023fwy}, the Bose-Hubbard model~\cite{saito2017solving}, the Schwinger model~\cite{Lin:2024eiz}, atomic systems and small molecules~\cite{han2019solving}, light nuclei~\cite{Adams:2020aax,Gnech:2021wfn,Gnech:2023prs}, and frustrated spins~\cite{choo2019two}.

Algorithms that make use of neural quantum states typically need some mechanism for efficiently sampling from the probability distribution $|\psi|^2$, where $\psi$ is the wavefunction defined by a neural network. For this reason several variations on the neural-network theme have been proposed, most notably autoregressive networks~\cite{PhysRevLett.124.020503}. The purpose of this paper is to propose and demonstrate an alternative approach based on normalizing flows.

Normalizing flows are a common machine-learning tool for approximating high-dimensional probability distributions in a way that makes sampling cheap without sacrificing the ability to evaluate the distribution $p(x)$ at a point. Normalizing flows have been applied to virtually every computational task for which sampling is a major cost, including (in nuclear physics) lattice field theory simulations~\cite{Kanwar:2020xzo,Boyda:2020hsi,Albergo:2021bna,Abbott:2022zhs,Abbott:2023thq} and Bayesian inference~\cite{Yamauchi:2023xrz}.

The connection between normalizing flows and neural quantum states is clear; the key difference is that a neural quantum state must represent phase information in addition to the probability distribution $|\psi|^2$. In this work we modify two common constructions of normalizing flows, RealNVP~\cite{dinh2016density} and continuous normalizing flows~\cite{chen2018neural}, to accomodate this phase information. The resulting object represents a quantum wavefunction $\psi(x)$ of a continuous variable, and simultaneously enables cheap parallel sampling from the probability distribution $|\psi|^2$.

The remainder of this paper is organized as follows. Section~\ref{sec:nf} introduces two modern architectures of normalizing flows and details their quantum generalizations, to be used to construct neural quantum states. In Section~\ref{sec:variational} we use these neural quantum states together with the variational method to approximate bound states of interacting particles in a harmonic trap, and in Section~\ref{sec:evolution} we demonstrate quantum time-evolution. Section~\ref{sec:error} describes the procedure for estimating systematic errors of the quantum evolution. Finally, in Section~\ref{sec:discussion} we conclude and discuss possible obstacles to the generalizability of the methods described in this work.

Results in this work are obtained using JAX~\cite{deepmind2020jax}, equinox~\cite{kidger2021equinox}, and diffrax~\cite{kidger2021on}.

\section{Normalizing flows and wavefunctions}\label{sec:nf}
A normalizing flow is an invertible map between a non-trivial ``target'' probability distribution and the normal distribution of the same dimension. Concretely, a normalizing flow $f:\mathbb R^N \rightarrow \mathbb R^N$ for a $N$-dimensional distribution $p(x)$ is defined to obey
\begin{equation}\label{eq:nf-def}
	p(x) \,dx = e^{-y^2 / 2} \,dy
	\text.
\end{equation}
Above $dx$ is a shorthand for the differential $N$-form $\bigwedge_{n=1}^N dx_n$. For computational purposes it is practical to eliminate the differentials in favor of a Jacobian matrix, yielding
\begin{equation}
	e^{-y^2 / 2} = p(x) \det \frac{\partial f}{\partial y}
	\text.
\end{equation}
A normalizing flow may be found for any probability distribution~\cite{villani2009optimal}; only in one dimension is the normalizing flow unique (up to parity).

From a computational standpoint, the virtue of a normalizing flow is that it allows efficient parallel sampling from the distribution $p$. Uncorrelated samples drawn from the normal distribution may be fed into $f(\cdot)$, yielding uncorrelated samples from $p(x)$. The advantage is somewhat larger in practice than in theory, as the evaluation of $f(\cdot)$ on many samples is vectorized on modern processors. Note also that by construction, the wavefunction is normalized such that $\int |\psi|^2 = 1$.

Exact normalizing flows are difficult to find and therefore of little use in numerical applications. Most frequently, one searches instead for approximate normalizing flows; that is, invertible maps $f(\cdot)$ for which the violation of Eq.~(\ref{eq:nf-def}) is small. The resulting algorithms therefore center around the minimization of some loss function measuring the distance between the desired distribution $p(x)$, and the distribution actually induced by the normalizing flow---Kullback-Liebler divergence and its variants are common choices.

In this section we will show how two modern constructions of normalizing flows can be modified to describe a quantum state $\psi(x)$ instead of just the probability distribution $(\psi^\dagger \psi)(x)$; in other words the normalizing flows are modified to include phase information. The desired wavefunction is not known in advance. Therefore, the central training step will involve minimizing not the distance between the represented wavefunction and some known distribution, but either the energy of the wavefunction (in Section~\ref{sec:variational}) or the violation of the time-dependent Schr\"odinger equation (in Section~\ref{sec:evolution}).

\subsection{Quantum continuous normalizing flows}\label{ssec:qcnf}
A continuous normalizing flow defines a map $y \rightarrow z$ via a differential equation
\begin{equation}
	\frac{d z}{d t} = F(z)
	\text.
\end{equation}
The initial condition is taken to be the input to the normalizing flow: $z(t=0) = y$. The output of the normalizing flow is taken from the evolution at time $t=1$. The function $F$ defining the differential equation is in this work defined according to a multi-layer perceptron (MLP)---that is, a dense neural network. This function has parameters $\alpha$, which then become parameters of the normalizing flow $f(y;\alpha)$ itself.

In practice, normalizing flows are difficult to use in many dimensions unless a fast algorithm is available for computing the induced probability distribution. A naive method for this computation requires computing the determinant of the Jacobian, which takes approximately cubic time in the number of degrees of freedom. With a continuous normalizing flow we can do far better. The logarithm of the determinant of the Jacobian evolves according to its own differential equation:
\begin{equation}
	\frac{d}{dt} \log \det J = \Tr H
	\text.
\end{equation}
Here $H$ refers to the Hessian of the neural vector field $F$, defined by
\begin{equation}
	H_{ij} = \frac{\partial}{\partial z_j} F_i(z)
	\text.
\end{equation}
In a system with $N$ degrees of freedom, the trace of the Hessian requires $N$ evaluations of the neural network defining $F$. For a dense network the resulting algorithm is still cubic time; a sparse network can reduce this to be as low as $O(N^2)$.

This standard construction provides only a probability, with no phase information. To accomodate a full wavefunction, we modify the neural network $F$ to have an additional output, giving the derivative of the phase:
\begin{equation}
	\left(\frac{d z}{dt},\frac{d\theta}{dt}  \right) = F(z)
\end{equation}
with $\theta(t=0)=0$. The wavefunction is constructed as
\begin{equation}
	\psi( x) = \left[\det\left(\frac{d x}{d y} \right)  \right]^{-1/2} \frac{1}{(2\pi)^{N/4}} e^{-\frac{ y\cdot y}{4} + i\theta(t=1)}\;\text.
\end{equation}

\subsection{QuantumNVP}

Our second construction of normalizing flows for neural quantum states is based on the widely used RealNVP architecture~\cite{dinh2016density}. First we summarize this architecture, and then describe its modification to permit description of quantum states in lieu of probability distributions (which we dub QuantumNVP).

RealNVP is centered around \emph{affine coupling layers}. An affine coupling layer is a map $y \rightarrow x$ defined by
\begin{equation}
	x_i = m_i y_i + (1-m)_i\left(y_i e^{s_i(y\odot m)} + t_i(y\odot m) \right)\;\text.
\end{equation}
In this map, $m \in \{0,1\}^N$ \textit{masks} the elements of $y$; the purpose of this will be clear when we discuss the computation of the Jacobian determinant. By u $\odot$ v we denote the element-wise product. The scaling and translation functions $s(\cdot)$ and $t(\cdot)$ are both functions $\mathbb R^{N} \rightarrow \mathbb R^{N}$ parameterized by neural networks; typically these are relatively shallow with between $0$ and $2$ nonlinear layers.

Because of the structure induced by the use of the mask $m$, the Jacobian matrix of this transformation is triangular. As a result, the determinant of the Jacobian of this transformation is simply the product of the diagonal elements of the Jacobian:
\begin{equation}\label{eq:affineJ}
	\log \det\left(\frac{\partial x}{\partial y}\right)
	=
	\sum_{n | m_n = 0} s_n(y \odot m)
	\text.
\end{equation}

A RealNVP normalizing flow consists of the composition of many affine coupling layers. It is important that the masks in the layers be different; in this work we divide the degrees of freedom into ``even'' and ''odd'' subsets, and each layer masks one of these two sets in an alternating fashion. The evaluation of the Jacobian determinant of a full RealNVP network may be performed cheaply, by multiplying the determinants of each constituent affine coupling layer.

Quantum NVP introduces the phase of the wavefunction by making the scaling function $s$ complex:
\begin{equation}
	 x_i = m_i y_i + (1-m)_i\Big[ y_i |s_i( y\odot m)|^2 + t_i( y\odot m) \Big]\text.
\end{equation}
Note that the scaling network that implements $s$ outputs $2N$ real numbers to accomodate phases; in the definition of $s(\cdot)$ we package these pairwise into $N$ complex numbers. The wavefunction is 
\begin{equation}
	\psi( x) = \left[\prod_{j=1}^N \left(s_j \right)^{1-m_j}\right]  \frac{1}{(2\pi)^{N/4}} e^{-\frac{ y\cdot y}{4}}
\end{equation}
with one quantum affine coupling layer. As before, the Jacobian of the map is the product of Eq.~(\ref{eq:affineJ}) over the affine layers, and the wavefunction becomes the Gaussian factor times the product of the total Jacobian factor.

\section{Variational states}\label{sec:variational}

\begin{figure}
	\includegraphics[width=0.95\linewidth]{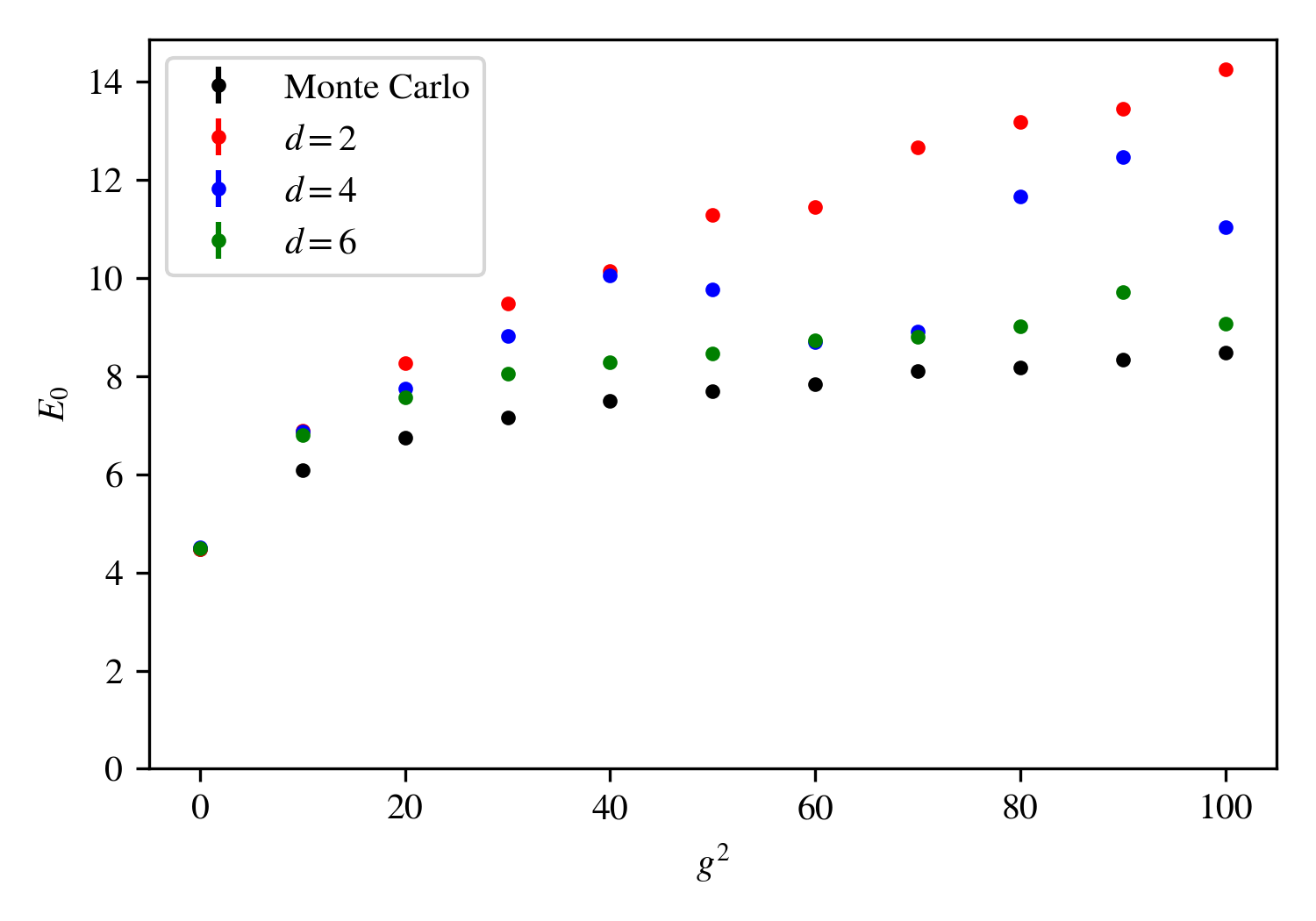}
    	\caption{Estimating the ground-state energy of the Hamiltonian in Eq.~(\ref{eq:Hg}) via QuantumNVP, as a function of coupling strength $g^2$. In black are shown Monte Carlo estimates of the ground-state energy. Other data points show $\langle H \rangle$ measured in optimized wavefunctions constructed from QuantumNVP with $d=2,4,6$ layers. Statistical errors on each measurement are shown, but too small to extend outside the datapoints. Estimates of the fluctuation from minimization to minimization are not provided.\label{fig:ground}}
\end{figure}

In this section we will use neural quantum states, specifically the QuantumNVP architecture described above, as a variational ansatz. For this experiment we consider three particles in three spatial dimensions, interacting via a Yukawa potential, in a harmonic trap. The Hamiltonian of this system is
\begin{align}\label{eq:Hg}
	\hat H_{\mathrm{trap}} &= \sum_n \left(\frac {\hat p_n^2}{2M} + \frac{M \omega^2}{2} \hat x_n^2\right) + \sum_{n<m} V_{nm}(\hat x)
	\text{,}\\
	& \text{where } V_{nm}( x) = \frac{g^2}{| x_n - x_m|}
	e^{-m |x_n - x_m|}\text.
\end{align}
The particles are taken to be distinguishable. For all simulations we adopt $M = \omega = 1$, standardizing the shape of the external potential, and $m=2$ setting the range of the potential. The interaction is repulsive when the coupling $g^2$ is positive; values of $g^2 \lesssim 10$ may be thought of as perturbative.

We use the QuantumNVP architecture with $d$ layers to define a family of wavefunctions $\psi(x;\alpha)$. The accompanying normalizing flows are denoted $x = f(y;\alpha)$, where $y$ are normally distributed variables in $\mathbb R^N$. Here $\alpha$ stands for all continuous parameters, on which gradient descent can be performed, in the neural network. The scaling functions $s(\cdot)$ in the (quantum) affine coupling layers are dense multi-layer perceptrons with one hidden/nonlinear layer using $\tanh$ as an activation function; the translation functions $t(\cdot)$ are taken to be linear.

To approximate the ground state, we define a loss function that evaluates $\langle \psi (\alpha)| H | \psi(\alpha) \rangle$:
\begin{equation}
	L_{\mathrm{ground}}[\alpha] = \frac{\int \frac 1 2 \left|\nabla \psi(x,\alpha)\right|^2 + |\psi(x,\alpha)|^2 V(x)\,dx}{\int |\psi(x,\alpha)|^2 \,dx}
\end{equation}
Note that we have integrated by parts in order to remove the Laplacian in favor of the (substantially cheaper) gradient squared. In performing gradient descent we approximate $L_{\mathrm{ground}}$ by pulling samples from the normal distribution and feeding them through the normalizing flow:
\begin{equation}
	L_{\mathrm{ground}}[\alpha] \approx \frac 1 B \sum_y \frac{\frac 1 2 |\nabla \psi|^2 + |\psi|^2 V(x) \,dx}{|\psi(f(y,\alpha),\alpha)|^2}
\end{equation}
A stochastic estimate of the gradient is obtained by automatic differentiation of this calculation. Gradient descent is performed with \texttt{Adam}~\cite{kingma2014adam} with standard hyperparameters and a learning rate of $3 \times 10^{-4}$. Each training step uses $B = 2^{10}$ samples, and we use $2^{15}$ samples for evaluating the Hamiltonian expectation value in the optimized state.

The weights defining the quantum affine couplings are drawn, at initialization, from the uniform distribution on $[-\frac{1}{dN},\frac{1}{dN}]$, where $d$ is the depth and $N$ is the number of degrees of freedom ($N=9$ in this problem). We train for $3 \times 10^4$ steps. To accelerate training we make use of layer normalization~\cite{ba2016layer}.

This system has no sign problem, and so the ground state can be efficiently examined via quantum Monte Carlo algorithms. For comparison purposes we perform a path-integral Monte Carlo calculation at an inverse temperature of $\beta = 10$, with an imaginary-time lattice spacing of $\Delta \tau = 10^{-1}$. Both parameters are chosen to make systematic errors negligible.

Figure~\ref{fig:ground} shows the energy of the system system as a function of the coupling constant $g^2$. A Monte Carlo calculation of the energy is performed for comparison purposes, along with minimizations of QuantumNVP wavefunctions with $d=2,4,6$ layers. All QuantumNVP calculations correctly reproduce the (Gaussian) noninteracting ground state; moreover all outperform first-order perturbation theory. The energies obtained by minimization are somewhat uneven from run to run---only one run is shown for each $(g^2,d)$ pair in this plot. At the greatest depth, $d=6$, the ground-state energy is approximated to within $10\%$.

It is worth noting that this exercise is not a strong test of the capabilities of QuantumNVP. The ground state of a bosonic Hamiltonian has no phases. As a result, defining a wavefunction by $\psi = \sqrt p$, with $p$ a probability distribution provided by RealNVP, would have sufficed.

\section{Time-evolution}\label{sec:evolution}

\begin{figure}
	\includegraphics[width=0.95\linewidth]{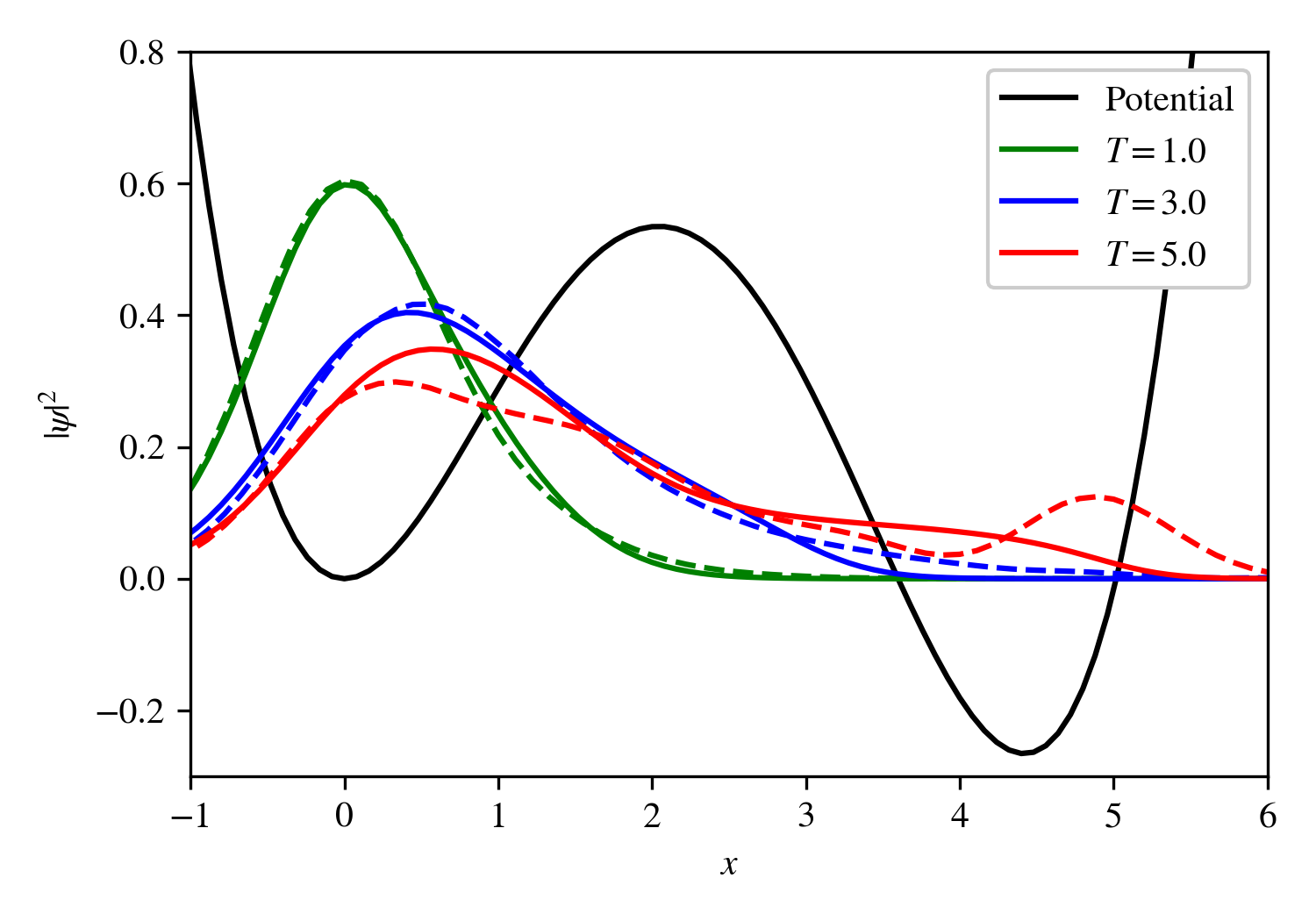}
	\caption{The differential probability of finding the particle at $x$, after time evolution. The Hamiltonian as in Eq.~(\ref{eq:Htunnel}), and the initial state is given by Eq.~(\ref{eq:psiunstable}). Results from the NFQS simulation are compared with the exact solution obtained by numerically solving Schr\"odinger's equation. \label{fig:tunneling}}
\end{figure}

The first method proposed for time-evolving neural quantum states~\cite{carleo2017solving} worked by considering the derivative of the wavefunction, $-i \hat H \psi$, as a vector in the space of wavefunctions. This vector was then projected onto the submanifold of wavefunctions representable by the chosen ansatz. This projection entails expressing the vector as a vector in the space of neural network weights. A single step of time-evolution in that direction is then taken, and the method is repeated until the desired amount of physical time has passed.

In this work we follow a later proposal~\cite{gutierrez2022real}. This method is detailed in this section (and error estimates are constructed in Section~\ref{sec:error} below). We will see that not only is the need for the pseudo-inverse removed, but the method allows us to take far larger time-steps without sacrificing fidelity.

We assume that a representation of the wavefunction at time $t_0$ has been found. In principle, this specifies a wavefunction at all later times by Schr\"odinger's equation:
\begin{equation}
	i \frac{\partial}{\partial t} \psi = \hat H \psi
	\text.
\end{equation}
We cannot hope to exactly construct $\psi(t)$ for two reasons; first because our computer must perform a finite number of operations and therefore cannot represent continuous time, and second because in general $\psi(t)$ for any fixed time will not be in our ansatz.

To address the first difficulty we replace Schr\"odinger's equation by its finite-differencing approximation:
\begin{equation}
	i \left[\psi(t+\Delta t) - \psi(t)\right] = \Delta t \hat H \frac{\psi(t+\Delta t) + \psi(t)}{2}
	\text.
\end{equation}
This defines a sequence of wavefunctions at times $t_0 + n \Delta t$ for integers $n$. The $(n+1)$th wavefunction is determined in principle from the $n$th.

We may now perform a single step of time-evolution, starting from the wavefunction at any time $t$, by minimizing the violation of this discretized form of Schr\"odinger's equation. Specifically we minimize the loss function
\begin{equation}\label{eq:loss-evolution}
	L_{\mathrm{evolution}}[\psi',\psi] = \int dx\, \Big|\frac{\psi' - \psi}{\Delta t} + i \frac{\hat H \psi + \hat H \psi'}{2}\Big|^2
	\text.
\end{equation}
The integral is computed by sampling from $|\psi|^2$ (using the corresponding normalizing flow) and reweighting.

We have used the lowest-order symmetric integration scheme; however this general algorithm is compatible with a wide variety of integration schemes. This was noticed already in~\cite{gutierrez2022real} where a symplectic integrator was proposed.

As a demonstration, we consider a particle tunneling from an unstable potential well to a nearby well of lower energy, through a barrier whose energy exceeds that of the particle. The Hamiltonian is
\begin{equation}\label{eq:Htunnel}
	H_{\mathrm{tun}} = \frac 1 2 p^2 +
	\frac 1 {2b^2} x^2 (x-b)^2 - \frac{a}{b^3} x^3
	\text.
\end{equation}
This form of the potential is chosen so that, no matter what values are selected for the shape parameters $a$ and $b$, the ``false vacuum'' corresponds to a quadratic potential normalized to $V_{\mathrm{false}} = \frac 1 2 x^2$. For this demonstration we select $a=0.25$ and $b=4.25$. With these parameters the local maximum of the potential, at $x \approx 2.0$, is $V(x) \approx 0.55$.

We use the quantum continuous normalizing flow architecture described in Section~\ref{ssec:qcnf}. We begin by preparing the ground state of the quadratic approximation to the potential around $x=0$, resulting in an NFQS approximation to
\begin{equation}\label{eq:psiunstable}
	\psi_{\mathrm{unstable}}(x) = \pi^{-1/4} e^{-x^2/2}
	\text.
\end{equation}
To observe tunneling, we then perform $50$ steps of real-time evolution under the tunneling Hamiltonian $\hat H_{\mathrm tun}$, each with a step size of $\Delta t = 0.1$. In Figure~\ref{fig:tunneling} we show the results of this simulation, specifically the probability density $|\psi(x)|^2$ at three different times $T$. The NFQS result is compared with nearly exact time-evolution obtained by discretizing Schr\"odinger's equation on a spatial lattice and numerically evolving as a coupled ODE. The wavefunctions are seen to be in good qualitative agreement. In the following section we will assign error bars to the measurement of certain observables, and show that good quantitative agreement is achieved as well.

\section{Error analysis}\label{sec:error}

Variational estimates of the ground state are famously plagued by the difficulty of quantifying the distance between the variational state and the true ground state. One may expand the ansatz (adding more layers and parameters to the neural network, in this case) and look for convergence, but strictly speaking there is no guarantee that ground state lies in or near the family of ans\"atze. Absent a complementary lower bound on the ground state energy (as may be provided by quantum mechanical bootstrap methods~\cite{Berenstein:2021dyf,Berenstein:2021loy,Lawrence:2022vsb}), variational estimates remain an uncontrolled approximation.

The same is not true for the time-evolution algorithm described in the previous section, as we will now show. We denote the true time-evolution $\psi(t)$, and the NQS approximation is written $\tilde \psi(t)$. We define an error function by
\begin{equation}
	E[\tilde \psi](t) = 1 - \Re \langle \tilde \psi | \psi\rangle
	\text.
\end{equation}
This measure of the error is valued on $[0,1]$ and tells us the extent to which the approximated evolution overlaps with the true evolution. In particular, defining an error vector by $|e\rangle \equiv |\psi\rangle - |\tilde \psi\rangle$, $E[\tilde \psi](t)$ is straightforwardly related to the norm:
\begin{equation}
	\big||e\rangle\big| = \sqrt{2 E[\tilde \psi](t)}
	\text.
\end{equation}
Note that this error function will be at its maximum when the two states are orthogonal, even if they are qualitatively similar.

By assumption ($\tilde\psi(0) = \psi(0)$), the error vanishes at $t=0$. It grows thereafter, and we may staightforwardly evaluate its growth:
\begin{equation}
	\frac{d}{dt} E[\tilde\psi](t) = \Re \langle \psi | \left[i H + \frac{d}{dt}\right] | \tilde \psi\rangle\text.
\end{equation}
This expression is not useful, as it contains reference to the true time-evolved wavefunction $\psi(t)$. However, noting that for any operator $\hat A$ we have the bound $\langle \psi | A | \tilde \psi\rangle \le |A | \tilde\psi\rangle|$, we can estimate the growth of the error by
\begin{align}
	\frac{d}{dt} E[\tilde\psi](t) &\le \left[\langle \chi | \chi\rangle \right]^{\frac 1 2} \\\nonumber
	&\text{where } |\chi\rangle \equiv \left(\frac{d}{dt} + i H\right) |\tilde\psi\rangle\text.
\end{align}
Here $\chi$ represents the difference between the time-derivative of $\tilde \psi$ and the true infinitesimal time evolution given by $-i H \chi$. In short, $|\chi\rangle = \frac{d}{dt} |e\rangle$.

Notice the correspondence with the loss function of Eq.~(\ref{eq:loss-evolution}) used to train the evolved wave function in the first place. That loss function was chosen precisely because it corresponds to the optimization of the overlap between the true time-evolution and the NQS approximation. As a result, we may estimate the error as
\begin{equation}
	E[\tilde\psi](T) \lesssim \int_0^{T} dt\,\sqrt{L_{\mathrm{evolution}}(t)}
	\text.
\end{equation}

The error estimates obtained from this method are rigorous as long as the $\Delta t \rightarrow 0$ limit and the limit of high statistics have been correctly taken. The overlap between the neural quantum state and the true time-evolved state is guaranteed to be at least as large as the bound obtained above.

Implicit in the above statement are three significant caveats. First, the error estimate does not account for any errors introduced in the initial state. Second, errors due to the discretization of time evolution have not been accounted for. These are insignificant for all results in this paper. However, optimizing the evolution algorithm requires understanding in detail the trade-offs in selecting $\Delta t$, which requires accounting for these errors correctly. Third and finally, although we can obtain a lower bound on the overlap of the true wavefunction with the numerically evolved one, there is no guarantee that this bound is close to being tight. In practice the bound is far from tight.

Typically the wavefunction itself is not of direct interest, and we instead wish to measure a time-evolved expectation value $\langle \mathcal O(t) \rangle$. The overlap bound above can be translated into an error estimate for this expectation value assuming that $\mathcal O$ has a known and finite operator norm $||\mathcal O||$:
\begin{equation}
	\langle \tilde \psi | \mathcal O | \tilde \psi \rangle
	- \langle \psi | \mathcal O | \psi \rangle
	< ||\mathcal O || \left(2\Big||e\rangle\Big| + \Big||e\rangle\Big|^2\right)
	\text.
\end{equation}
This bound uses only quantities that are already computed in the course of evolving the NQS. Since at each time step we perform gradient descent until the loss drops below some fixed threshold $L_{\mathrm{th}}$, the error estimate on an evaluated expectation value is proportional to $T L_{\mathrm{th}}$.

The bound we use on the norm of the error vector $|e\rangle$ is very loose in practice; consequently, this bound on the error in $\langle \mathcal O \rangle$ is similarly loose. We can improve the situation at the cost of some rigor. The infinitesimal error vectors $|\chi\rangle$ will typically not add coherently. As a result, $|e\rangle$ does not grow linearly with time, but instead performs a random walk. We no longer have a strict bound on the growth of this vector, but we expect it to behave as
\begin{equation}
	\Big||e\rangle\Big| \sim (\Delta t) L_{\mathrm{th}} T^{-1/2}
	\text.
\end{equation}

\begin{figure}
	\centering
	\includegraphics[width=0.95\linewidth]{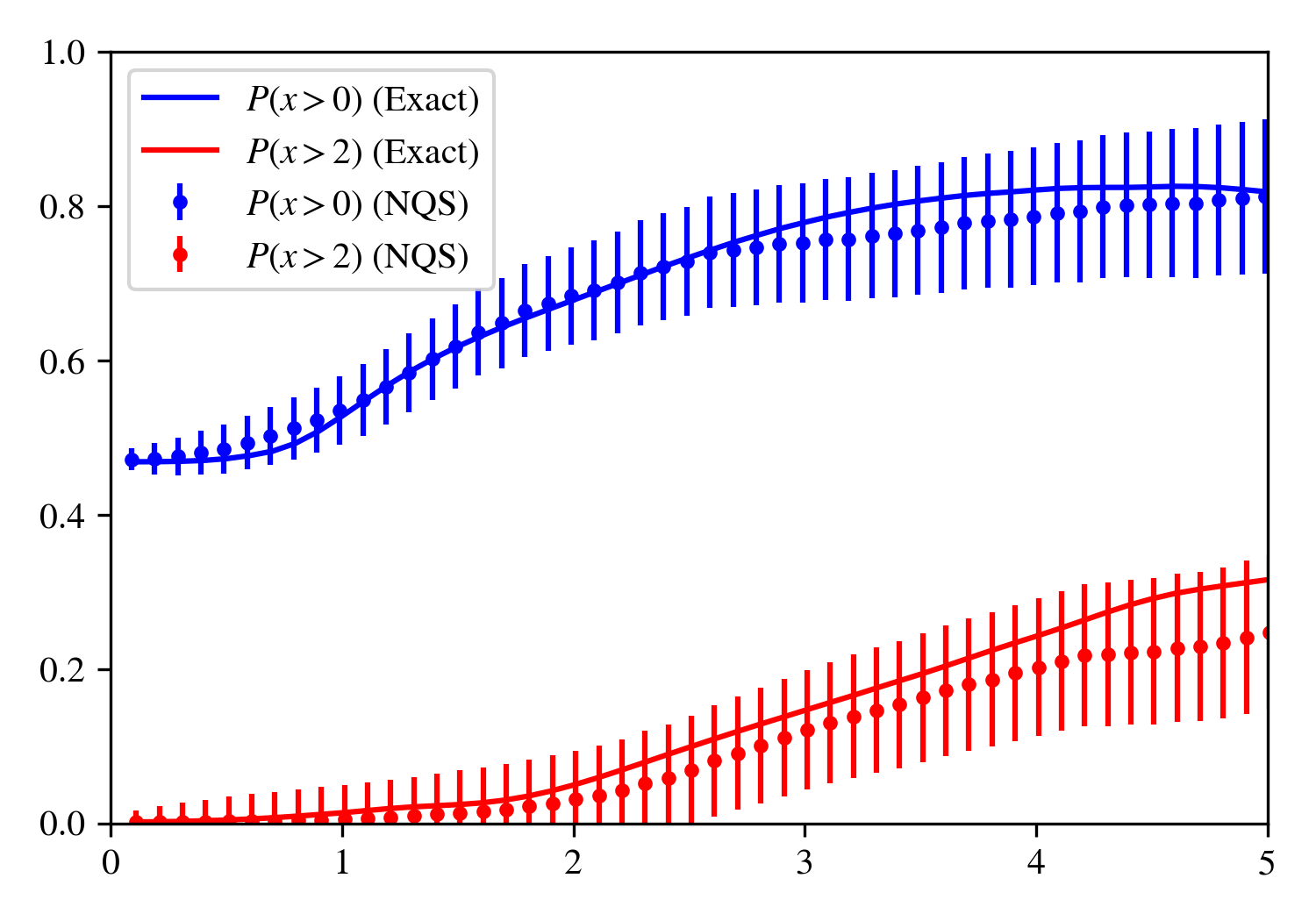}
	\caption{The probability of finding the particle in various regions as a function of time. Exact results are obtained by numerically solving the PDE as obtained, and neural quantum state results are obtained from the data used to generate Figure~\ref{fig:tunneling}, with error bars computed as described in Section~\ref{sec:error}.\label{fig:error}}
\end{figure}

As a simple example, we apply this bound to the tunneling simulation of the previous section. Figure~\ref{fig:tunneling} showed the time-evolved wavefunctions. We cannot obtain meaningful pointwise error estimates for the wavefunction itself, as the wavefunction at a single point may change arbitrarily without affecting the error $E$ defined in terms of the inner product of states. However, we can define an operator
\begin{equation}
	\hat\Theta = \theta(\hat x-x_0)
\end{equation}
such that $\langle \hat \Theta \rangle$ measures the probability of finding the particle in the lower well. Here $x_0$ is taken to be the position of the local maximum of the potential between the two wells. The expectation value of this operator begins at approximately $0$ in the initial state, and slowly rises as the particle tunnels out of its metastable state. Because this operator has bounded eigenvalues, we are able to estimate its errors as above. The results are shown in Figure~\ref{fig:error}.

\section{Discussion}\label{sec:discussion}

We have presented two architectures of neural quantum states, both based in normalizing flows. The advantage to these architectures comes from the ease of sampling with respect to the wavefunction-induced probability $|\psi|^2$. These architectures are usable for both ground-state preparation (Section~\ref{sec:variational}) and real-time evolution (Section~\ref{sec:evolution}), as with other types of neural quantum states.

Although variational methods for ground-state energy estimation do not come with rigorous error bars, the systematic errors in real-time evolution of neural quantum states can be quantified---or at least, rigorously upper bounded. We obtained such bounds in Section~\ref{sec:error}. The bounds obtained are clearly quite loose. It remains for future work to investigate whether these bounds can be made tighter without incurring unreasonable computation cost.

The constructions of neural quantum states in this paper share one apparent shortcoming: they have difficulty representing winding number. A generic wavefunction in two or more dimensions will have zeros, living on surfaces of codimension $2$. About such a surface there may be a winding number, which is to say that the wavefunction locally takes the form $\psi(\theta) \sim e^{n i \theta}$. The key property of such a wavefunction is that, due to this winding number, the wavefunction cannot be well approximated by any wavefunction that lacks zeros (and thus lacks a winding number).

Neither continuous normalizing flows nor QuantumNVP is able to create such a winding number where none previously existed. In particular, when the starting distribution is a simple Gaussian, the final wavefunction will have no zeros that have winding numbers. In this work we disregarded this issue and were able to obtain good results anyway. In the case of Section~\ref{sec:variational} this is because bosonic ground states lack zeros; in the case of the time-evolution performed in Section~\ref{sec:evolution} this is because winding numbers are absent in one dimension. In general we expect it to be necessary to have an ansatz supporting zeros: the first excited state of the hydrogen atom notably has a winding number. One possibility is to begin the normalizing flow not with a Gaussian but with a wavefunction that already has winding numbers present. We leave this question to future work.

Finally, we note that realistic quantum simulations of first-quantized systems require subsets of particles to have either symmetric or antisymmetric statistics imposed. Other constructions of neural quantum states have been suitably modified to support indistinguishable particles in this sense (see for example~\cite{Bedaque:2023udu} for bosons and FermiNet~\cite{PhysRevResearch.2.033429} for fermions). This has yet to be done for quantum states constructed from normalizing flows.

\acknowledgments
S.L.~and A.S.~were supported at the beginning of this work by the U.S.~Department of Energy under Contract No.~DE-SC0017905. S.L.~is subsequently supported by a Richard P.~Feynman fellowship from the LANL LDRD program. A.S.~is subsequently supported by NSF award PHY-2209590. Y.Y.~is supported by the INT's U.S. Department of Energy grant No.~DE-FG02-00ER41132. Los Alamos National Laboratory is operated by Triad National Security, LLC, for the National Nuclear Security Administration of U.S. Department of Energy (Contract No.~89233218CNA000001).

\bibliographystyle{apsrev4-2}
\bibliography{refs}

\end{document}